\documentclass[11pt,a4paper]{article}
\usepackage{jcappub}
\usepackage{ifthen}
\usepackage{epsfig}
\usepackage{epstopdf}

\usepackage{color}

\usepackage[T1]{fontenc}

\newcommand{\dd}{\textrm{d}}
\newcommand{\n}{\ensuremath{\mathbf{n}}}

\title{CMB Aberration and Doppler Effects as a Source of Hemispherical Asymmetries}

\def\bar{Departament de F\'isica Fondamental i Institut de Ci\'encies del Cosmos, Universitat de Barcelona, Mart\'i i Franqu\'es 1, E-08028 Barcelona, Spain}
\def\fer{Dip. di Fisica, Università di Ferrara and INFN Sez. di Ferrara, Via Saragat 1, I-44100 Ferrara, Italy}
\def\rio{Instituto de Física, Universidade Federal do Rio de Janeiro, CEP, 21941-972, Rio de Janeiro, RJ, Brazil}
\def\goe{Institut f{\"u}r Theoretische Physik, Friedrich-Hund-Platz 1, 37077 G{\"o}ttingen, Germany}

\author[a,b]{Alessio Notari,}
\author[c]{Miguel Quartin}
\author[d]{and Riccardo Catena}
\affiliation[a]{\bar}
\affiliation[b]{\fer}
\affiliation[c]{\rio}
\affiliation[d]{\goe}

\abstract{
    Our peculiar motion with respect to the CMB rest frame represents a preferred direction in the observed CMB sky since it induces an apparent deflection of the observed CMB photons (aberration) and a shift in their frequency (Doppler). Both effects distort the multipoles $a_{\ell m}$'s at {\emph all} $\ell$'s.  Such effects are real as it has been recently measured for the first time by Planck according to what was forecast in some recent papers.     However, the common lore when estimating a power spectrum from CMB  is to consider that Doppler affects only the $\ell=1$ multipole, neglecting any other corrections. In this work we use simulations of the CMB sky in a boosted frame with a peculiar velocity $\beta\equiv v/c =1.23\times 10^{-3}$ in order to assess the impact of such effect on power spectrum estimations in different regions of the sky.  We show that the boost induces a north-south asymmetry in the power spectrum which is highly significant and non-negligible, of about $(0.58 \pm 0.10)\%$ for half-sky cuts when going up to $\ell \approx 2500$.     We suggest that these effects are relevant and may account for some of the north-south asymmetries seen in the Planck data, being especially important at small scales. Finally we analyze the particular case of the ACT experiment, which observed only a small fraction of the sky and show that it suffers a bias of about $1\%$ on the power spectrum and of similar size on some cosmological parameters: for example the position of the peaks shifts by $0.5\%$ and the overall amplitude of the spectrum is about $0.4\%$ lower than a full-sky case.}


\begin{document}
\maketitle

\section{Introduction}
The Cosmic Microwave Background is used as a fundamental tool to test cosmological models and to quantitatively estimate the parameters of such models. This is usually done by extracting the temperature  and the polarization power spectra from the maps and by fitting them with a $\Lambda$CDM background model with an almost scale-invariant gaussian spectrum of density perturbations. However, if we observe the CMB with a velocity $\beta\equiv v/c$ relative to such background, the image undergoes distortions due to the Doppler and aberration effects and the correct procedure should be to first transform the image in the CMB rest frame and then analyze the data. Unfortunately such procedure is not performed by any of the current experimental analysis.

Such boost in a direction defined by the unit vector $\hat{\mathbf{z}}$  distorts a primordial temperature map $T(\hat{\mathbf{n}})$ by a Doppler factor and by changing the arrival direction~\cite{Challinor:2002zh,Amendola:2010ty}:
\begin{equation}
    T^{\prime}(\hat{\mathbf{n}}') =  \gamma(1+\beta\hat{\mathbf{n}}\cdot\hat{\mathbf{z}}) T(\hat{\mathbf{n}}) \,,
    \label{boost}
\end{equation}
where  $(\hat{\n}' - \hat{\n}) \cdot \hat{\mathbf{z}} \;= \beta \sin^2\theta /(1+\beta  \cos\theta)$, $\,\cos\theta = \hat{\bf n}\cdot\hat{\bf z}\,$ and $\,\gamma\equiv1/\sqrt{1-\beta^2}$.
Such distortion introduces correlations between different $a_{\ell m}$'s (the coefficients of the spherical harmonics decomposition), both \emph{diagonally} (i.e., between same $\ell$'s) and \emph{off-diagonally} (between different $\ell$'s)~\cite{Challinor:2002zh}:
\begin{equation}
    \textstyle
    a^{[\rm Boosted]}_{\ell m}\; =\; \sum_{\ell'} K_{\ell' \, \ell\, m} \, a_{\ell' m}^{[\rm Primordial]}\,.
    \label{matrixalm}
\end{equation}
The only effect which is usually taken into account is the large $\ell=1$ dipole, due to the matrix element $K_{0 1 0}$, which is used to infer our velocity $\beta$ \cite{Lineweaver:1996xa}. Clearly however there is much more information in the other matrix elements.
The off-diagonal correlation was shown in~\cite{Kosowsky:2010jm,Amendola:2010ty,Notari:2011sb} to be measurable by Planck as an alternative method to measure $\beta$ and in fact the Planck collaboration itself~\cite{Aghanim:2013suk} recently published  for the first time a detection of $\beta$ through CMB aberration and Doppler using $500<\ell\lesssim2000$. This provides an alternative measurement of $\beta$ with respect to the usual $\ell=1$ Doppler effect.


In this work we assess whether ignoring such frame effect may also induce a significant spurious directional dependence in the CMB power spectrum. In particular we analyze the pseudo-$C_{\ell}$ reconstructed in different regions of the sky by performing simulations in the case in which a boost effect is present or not. The pseudo-$C_{\ell}$'s (dubbed simply $\tilde{C}_{\ell}$'s) are defined simply as $\tilde{C}_{\ell}\equiv \sum_m |a_{\ell m}|^2/(2\ell+1)$ where the $a_{\ell m}$ are taken from the masked image. In the absence of a mask or instrumental noise, they coincide with the real $C_{\ell}$'s. If a mask is present, however, one must de-convolve the mask
\begin{equation}
C_{\ell} \;=\; \sum_{\ell'} M^{-1}{}_{\ell \ell'} \,  \tilde{C}_{\ell'}
\end{equation}
where the $M_{\ell \ell'}$ matrix is the mode-mode coupling kernel~\cite{Hivon:2001jp} which has to be inverted.

We proceed here as in \cite{Catena:2012hq}: we simulate maps of the CMB sky and we directly apply on the maps the boost transformation before extracting the $a_{\ell m}$'s, therefore bypassing the need of computing the  mixing coefficients. Then we extract the $a_{\ell m}$'s and the $C_{\ell}$'s and show that an asymmetry is visible. We also roughly quantify the amount of asymmetry through a coefficient which is built to estimate the change of the overall amplitude of the power spectrum. We apply all this procedure to temperature maps for an ideal experiment (neglecting instrumental noise) for 3 different sky-cuts: 2 composed of antipodal discs with measurable fraction of the sky $f_{\rm sky} = 0.88$ and $0.29$, and one with the exact observable area of the Atacama Cosmology Telescope (ACT)~\cite{Sievers:2013wk}. We compute our results up to a maximum multipole $\ell_{\rm max}=3000$. Our results can thus easily represent a Planck-like case (almost all-sky and $\ell_{\rm max}\simeq2500$), a WMAP-like case (almost all-sky and $\ell_{\rm max}\simeq900$) and some other high-precision small-area surveys, with $\ell_{\rm max}\simeq3000$ and similar $f_{\rm sky}$ as SPT~\cite{Keisler:2011aw}. For ACT in particular, we show the effect on the actual observed sky area, which is composed two thin regions called South and Equatorial stripes~\cite{Dunkley:2013vu}.

We apply our procedure to extract $\tilde{C}_{\ell}$'s from different regions of the sky in simulations performed with the HEALPix package\footnote{\url{http://healpix.sourceforge.net/}} in a modified version which allows boosts (originally made in~\cite{Catena:2012hq}), with $N_{\rm side}=4096$ and $\ell_{\rm max}=3000$. Such a numerical boost procedure has been tested with Bessel fitting functions which reproduce the $K_{\ell' \ell m}$ with high precision  \cite{Notari:2011sb,Catena:2012hq}.\footnote{We have checked up to $\ell_{\rm max}=2500$ that boosting a $N_{\rm side}=4096$ map with $\beta=0.00123$ and boosting again with $-\beta$ gives back the original $a_{\ell m}$'s with very high accuracy (and even more so for the $C_{\ell}$'s), differently from what was obtained in \cite{Yoho:2012am}.} In all cases we apply the masks smoothed with a beam of $10$ arcmin FWHM in order to minimize contamination at higher $\ell$.  For the case of ACT stripes, we go further and compute the $C_\ell$'s obtained from the $\tilde{C}_{\ell}$'s and the mask through the mode-mode coupling kernel, ignoring instrumental noise. Since in our analysis we will bin the $C_\ell$'s in $50$-$\ell$ bins it is better to work with $D_{\ell} \equiv \ell (\ell+1) C_{\ell}  / (2\pi)$ instead of the $C_\ell$'s and $\tilde{C}_\ell$'s because the $D_{\ell}$'s change more slowly as a function of $\ell$.

It is worth to note that traditionally in CMB analysis from the measured intensity map one does not compute the exact equivalent thermodynamical temperature map through the relation~\cite{Amendola:2010ty}
\begin{equation}
    T = \frac{\nu}{\log\left(1+\frac{\nu^3}{I(\nu)}\right)} \,,
    \label{eq:temp-intens}
\end{equation}
and Planck is no exception. Instead, one estimates the temperature through a linear order transformation assuming a black-body radiation spectrum. This procedure is fine when one is concerned  with first-order perturbation quantities, but leads to distortions at higher-order quantities, such as the aberration effect. In practice, for aberration this can be circumvented by computing the corrections explicitly, as was done in~\cite{Aghanim:2013suk}. These corrections appear as frequency dependent ``boost factors''. Here, since we assume to be working with the exact thermodynamical temperature, we can neglect these corrections.

\section{Power spectra North-South asymmetry}\label{sec:asymmetry}

\begin{figure}[t]
\begin{minipage}[c]{0.48\linewidth}
    \centering
    \includegraphics[width=\textwidth]{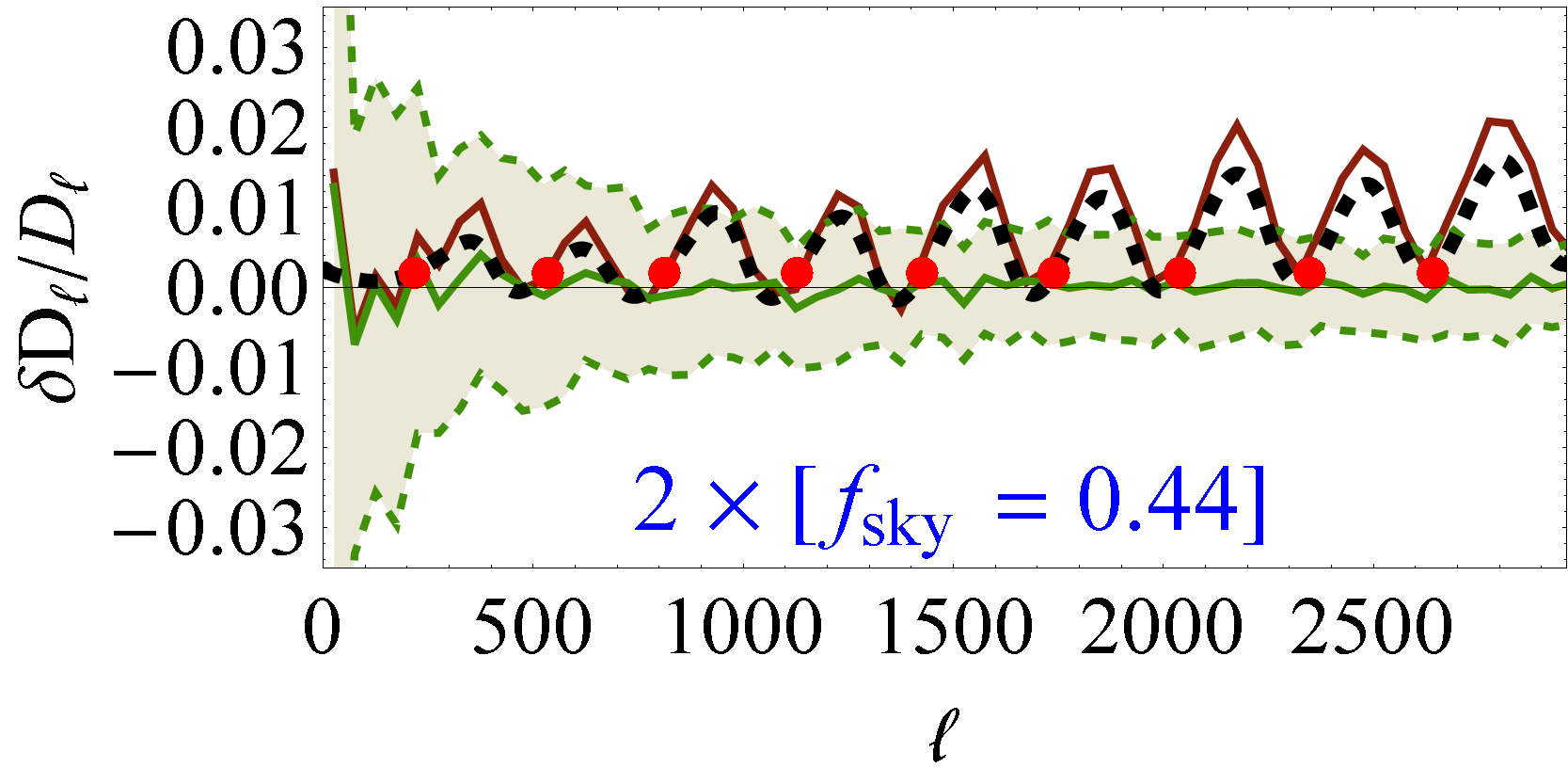}
    \includegraphics[width=\textwidth]{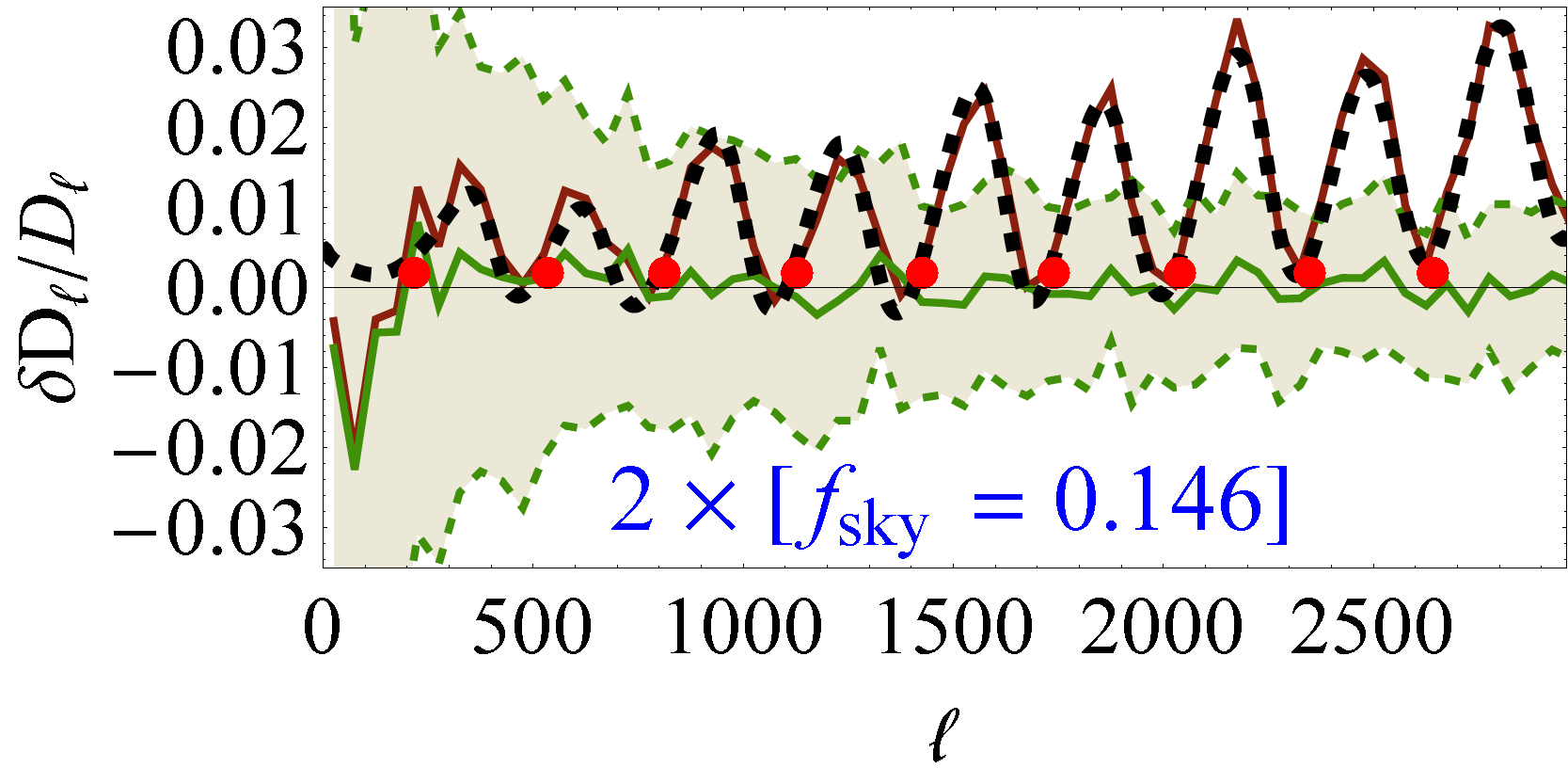}
\end{minipage}
\begin{minipage}[c]{0.54\linewidth}
    \includegraphics[width=\textwidth]{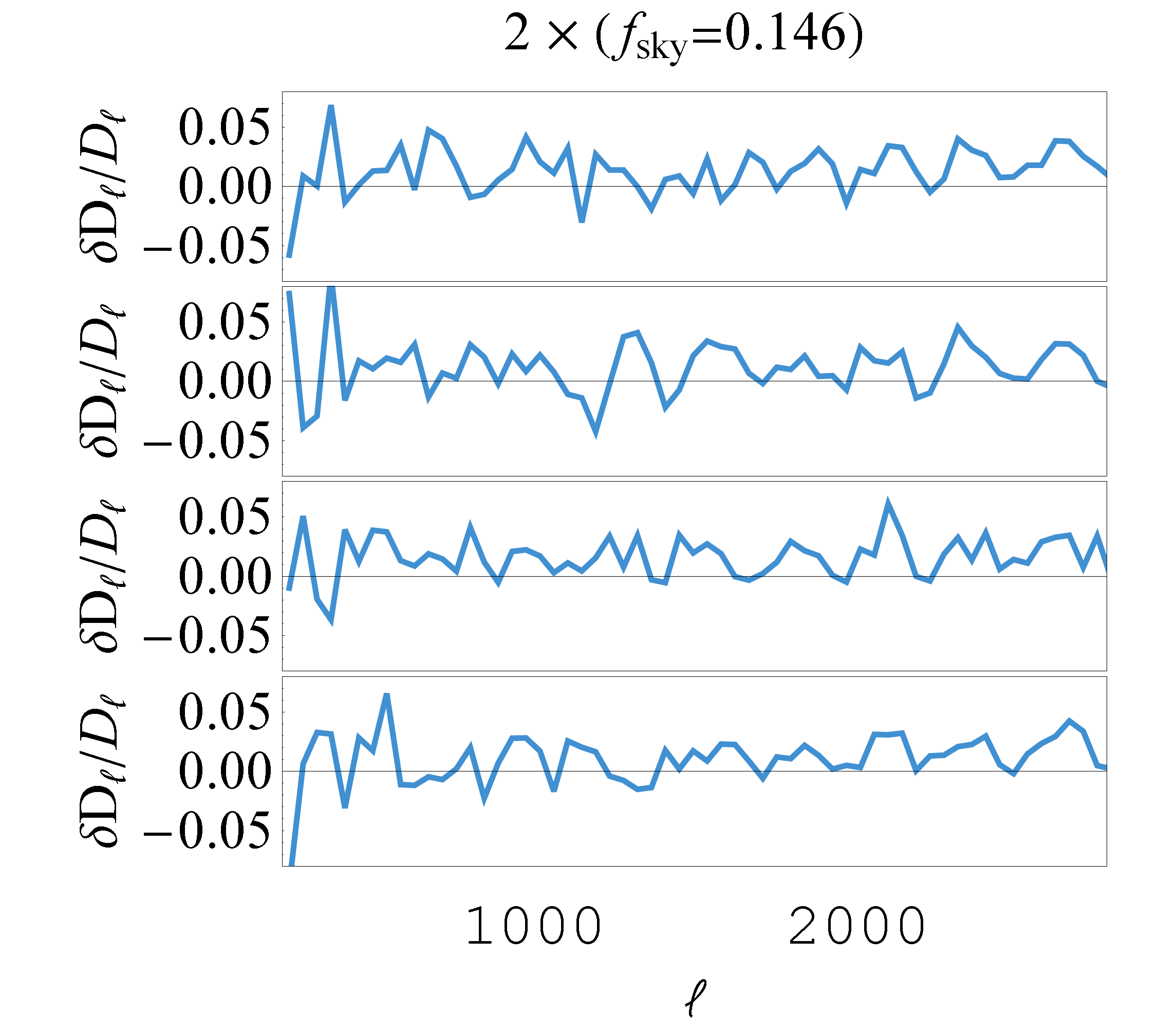}
\end{minipage}
    \caption{Relative difference between $D_{\ell}$'s in two opposite discs of the sky  centered on the dipole direction  as a function of the multipole $\ell$.  The brown (green) line shows the mean spectrum over 50 boosted (unboosted) simulations, binned in 50-$\ell$ bins. The black dashed curves are the analytical estimate of Eqs.~\eqref{eq:dAoverAest}--\eqref{averagebeta}. The green-shaded area shows the binned $1\sigma$ band around the unboosted mean. \emph{[Left Top]}: two {\emph halves} of the sky ($f_{\rm sky} = 0.44$, after removing a band around the galaxy).
    \emph{[Left Bottom]}: two antipodal discs of $90^\circ$ diameter ($f_{\rm sky} = 0.146$,  similarly to the recent Planck collaboration asymmetry analysis~\cite{Ade:2013nlj}). \emph{[Right]}: 4 random boosted realizations. Here, and in the next figures, the red dots show the position of the first 9 peaks of the fiducial CMB spectrum.}
    \label{compare5}
\end{figure}

We apply a boost along the north pole direction of $\beta=0.00123$ and compare two regions in two opposite directions along such boost (dubbed ``North'' and ``South''). We performed 50 different simulations as random realizations, labelled by the random seed used as an input in HEALPix, of a fiducial cosmological model similar to the WMAP 9-year best fit. We find that there is a systematic asymmetry between power in the two opposite directions in the case  with boost as opposed to the case without boost. We show the difference in power spectra in Fig.~\ref{compare5}. This should be compared with the experimental results presented in~\cite{Ade:2013nlj} (version v1, section 5.5.1 and Fig.~28), from which we can see that  the real data has systematically more power in one hemisphere than the other at a level of a few percent,  although it is difficult to make a more precise estimate on the significance and scale of the effect based on the results presented at that paper.

Given a set of $D_{\ell}$'s it is possible to have a rough estimate on the size of the effect on a cosmological parameter, by considering \cite{Catena:2012hq} an idealized case  in which the CMB depends multiplicatively on a single amplitude parameter, which we call $A$, so that the $\chi^2$ is given by:
\begin{equation}
    \chi^2(A)=\sum_\ell \frac{(D^{\rm exp}_{\ell}-A \hat{D}^{\rm th}_{\ell})^2}{\sigma^2_\ell}\,,
    \label{chi2}
\end{equation}
where $\hat{D}^{\rm th}_{\ell}$ is the theoretical spectrum when $A=1$, $D_{\ell}^{\rm exp}$ are the observed values in one region of the sky and where $\sigma^2_\ell=D_{\ell}^2 \;2/(2\ell+1)$ is the cosmic variance, ignoring any noise.
The best fit value $A_{\rm bf}$ for $A$ is obtained when $\partial (\chi^2)/\partial A=0$ which gives:
\begin{equation}
    A_{\rm bf}=\sum_\ell \frac{D^{\rm exp}_{\ell} \hat{D}^{\rm th}_{\ell}}{\sigma^2_\ell} \bigg / \sum_{\ell} \frac{(\hat{D}^{\rm th}_{\ell})^2}{\sigma_\ell^2} \,,
\end{equation}
The estimate of the relative difference 
between the best-fit values in two regions $N$ (North) and $S$ (South) of the sky with observed spectra $D^{\rm exp,\, N}_{\ell} $ and $D^{\rm exp ,\, S}_{\ell} $ is therefore given by:
\begin{equation}
    \left<\frac{\delta A}{A}\right> \,\equiv\, 2\frac{A^N_{\rm bf}-A^S_{\rm bf}}{A_{\rm bf}^N+A_{\rm bf}^S}
     \,\simeq\, \frac{ \sum_\ell (2\ell+1) \delta D_{\ell}/D_{\ell} }{    \sum_{\ell} \left(2\ell+1  \right)}\,,
    \label{biasAest}
\end{equation}
where $\delta D_{\ell} \equiv D^{\rm exp,\, N}_{\ell}-D^{\rm exp,\, S}_{\ell}$ and we have approximated $D_\ell^{\rm th}$ by the average  $(D^{\rm exp,\, N}_{\ell}+D^{\rm exp,\, S}_{\ell})/2\,$. We therefore estimate\footnote{Such an estimator has been used also by \cite{Dai:2013kfa} to quantify hemispherical asymmetries.} $\delta A/A$ for our simulations summing up to a certain  $\ell_{max}$. We will compute the estimator directly using the pseudo-$D_{\ell}$'s given as an output by HEALPix in~(\ref{biasAest}). For a sufficiently large patch of the sky  the offset between pseudo-$D_{\ell}$ and $D_{\ell}$ is simply an overall factor given by the sky fraction $f_{\rm sky}$ so (\ref{biasAest}) is still a good estimator of the difference in  amplitude of the two hemispheres. The same equation can also be used for small sky patches but in that case in principle one should use the reconstructed $D_{\ell}$ instead of the pseudo-$D_{\ell}$, otherwise (\ref{chi2}) should have a full covariance matrix (nevertheless as we will show these corrections in the case of ACT are small). The typical values we obtain for $\delta A/A$  are small and centered around zero for the case without aberration  while they represent an important effect of order $1\%$ when aberration is present.

It is interesting to note that $\delta D_{\ell}/D_{\ell}$ in~\eqref{biasAest} can be well approximated by a simple analytical estimate. For a small circle around the pole defined by the boost, the Doppler effect induces a shift of $\pm\beta$ in temperature amplitude (and thus a $4\beta$ north-south discrepancy in power), whereas aberration can be understood as a shift in $\ell$-space of $\sim \pm \beta \ell$~\cite{Burles:2006xf}. We  can thus estimate $\delta A/A$ by approximating the aberration in a given hemisphere as a simple change in $D_\ell$ given by $\Delta D_\ell = (\dd D_\ell / \dd \ell) \,\Delta \ell \sim \pm (D_{\ell+1}-D_\ell) \Delta \ell$. The North--South difference is thus $-2(D_{\ell+1}-D_\ell) \Delta \ell$ and we get
\begin{equation}\label{eq:dAoverAest}
    \frac{\delta D_\ell}{D_\ell}  \;\simeq\;  4 \overline{\beta} \,+\, 2 \overline{\beta} \, \ell \, \bigg(1-\frac{D_{\ell+1}^{\rm th}}{D_{\ell}^{\rm th}}\bigg)\,,
\end{equation}
where $\overline{\beta}=\beta$ for a small disc around  the pole. For a larger disc or for a generic region eq.~\ref{eq:dAoverAest} still holds if $\overline{\beta}$ is an average
\begin{equation}
\overline{\beta}=\beta \int_{R} d\Omega \cos(\gamma) \label{averagebeta}
\end{equation}
where $R$ is the region of interest and $\gamma$ is the angle relative to the boost direction.


\begin{figure}[t]
    \centering \includegraphics[width=.62\columnwidth]{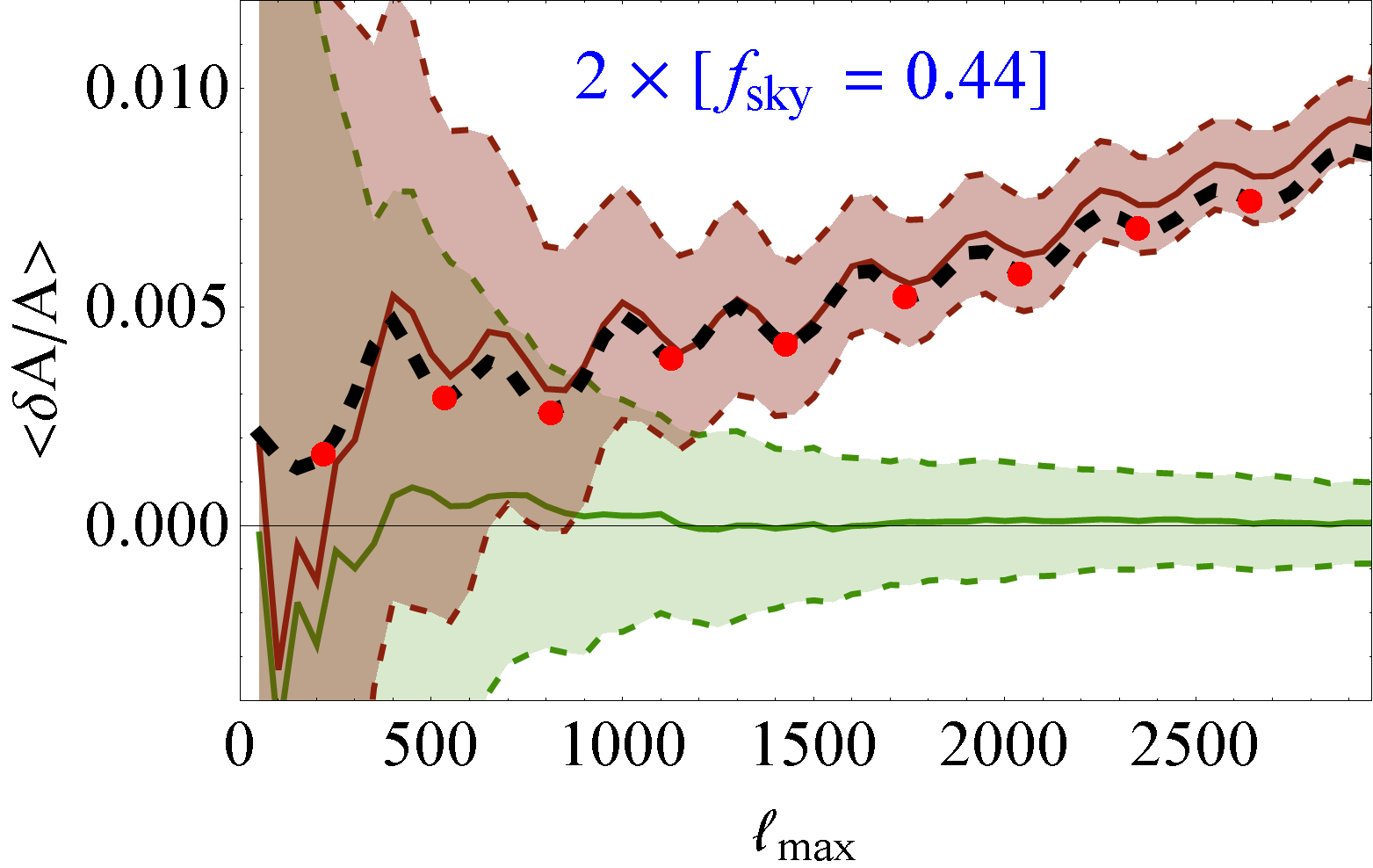}
    \includegraphics[width=.62\columnwidth]{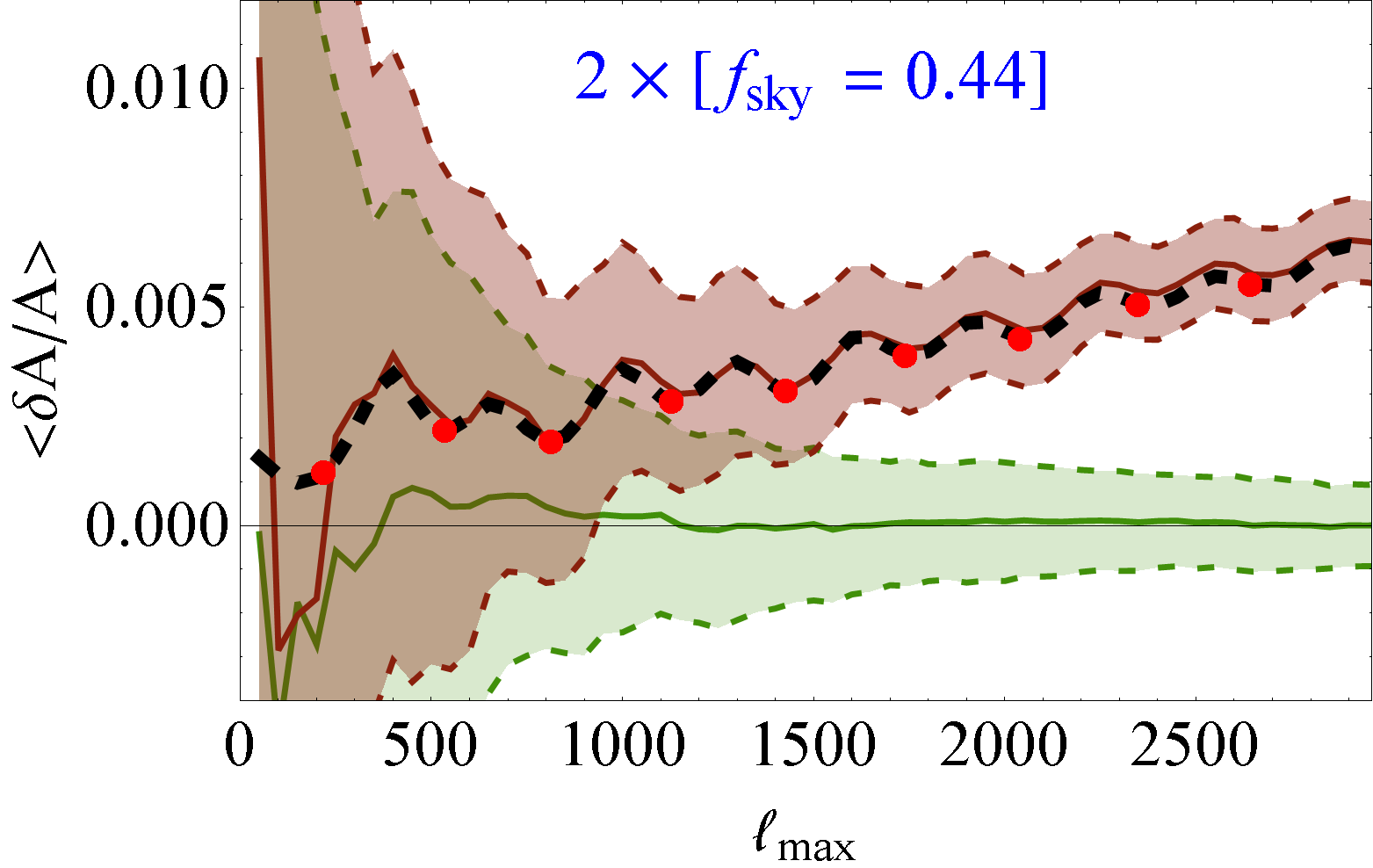}
    \caption{Average and $1\sigma$ bands for 50 realizations of the North-South asymmetry parameter $\delta A/A$ of Eq. (\ref{biasAest}). Brown curves are made on boosted maps, green on non-boosted ones and the thick black dashed curve is the analytical estimate of Eqs.~\eqref{biasAest}--\eqref{averagebeta}. The difference is computed using  pseudo-$D_{\ell}$'s in two {\it halves} of the sky ($f_{\rm sky} = 0.44$). \emph{[Top]} boost towards the north pole. \emph{[Bottom]}: boost along the actual direction given by the measured $\ell=1$ CMB dipole,  [$(l,b)=(264^\circ,48^\circ)$] in galactic coordinates. }
    \label{asymmetryell}
\end{figure}

\begin{figure}[t]
    \centering
    \includegraphics[width=.65\columnwidth]{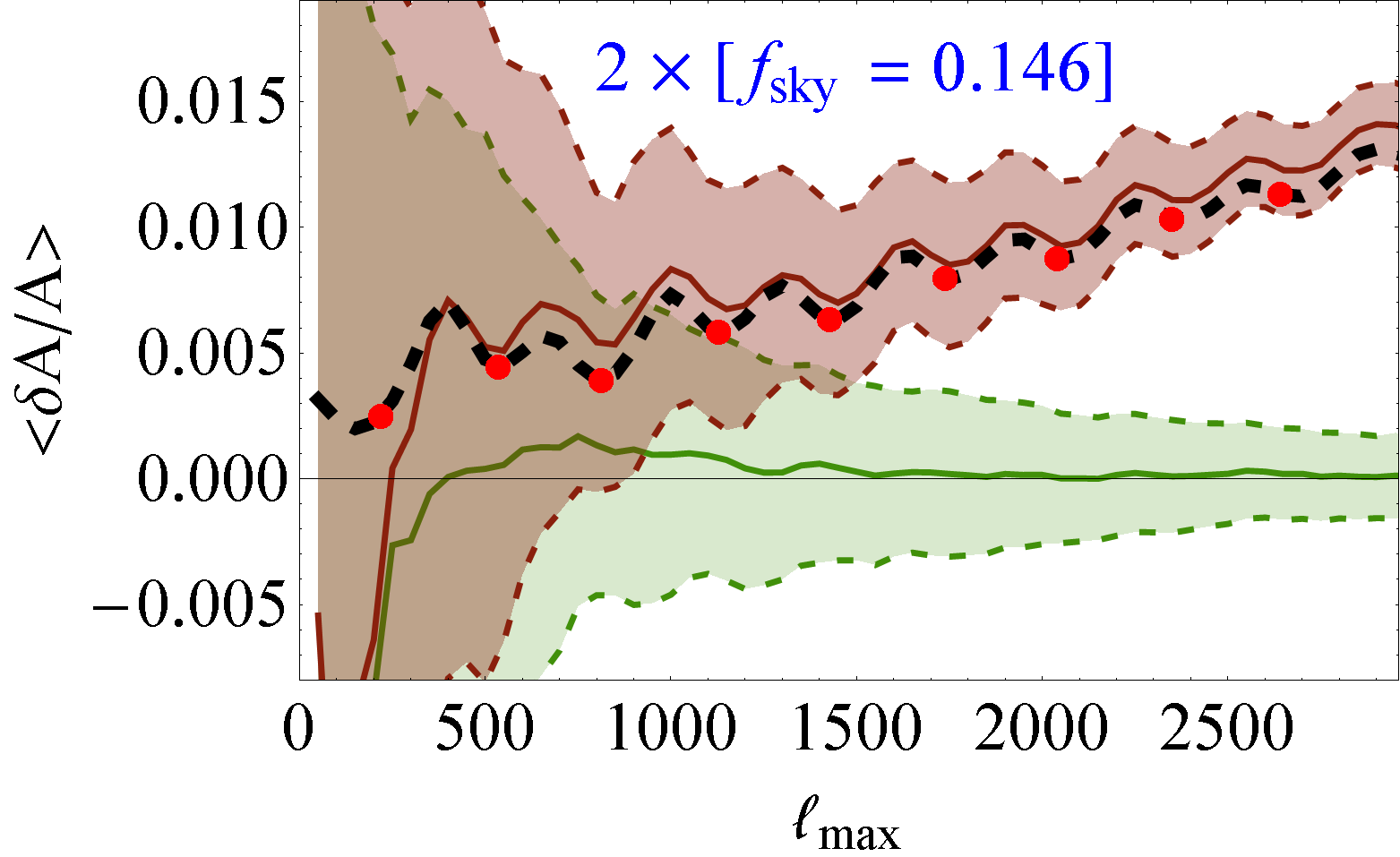}\quad {}
    \caption{Same as Fig.~\ref{asymmetryell} for a smaller patch of the sky, consisting of 2 antipodal discs of $90^\circ$ diameter ($f_{\rm sky} = 0.146$, cf. with Fig. 28 in v1 of~\cite{Ade:2013nlj}).   Note that although the discrepancy is larger for smaller $f_{\rm sky}$, this is over-compensated by the increase of variance among different realizations. Nevertheless the observed CMB is one particular realization and so the difference on a parameter estimation may well be large (of order a few percent).  Such results, when divided by 2, also represents the bias of a single disc experiment on the amplitude of the power spectrum and presumably on other cosmological parameters. }
  \label{asymmetryell2}
\end{figure}

Figs.~\ref{asymmetryell} and~\ref{asymmetryell2} show the result for mean and standard deviation of the asymmetry parameter as a function of $\ell_{\rm max}$ for our 50 simulations. Such an asymmetry is definitely detectable: at $\ell_{\rm max}\approx 2000$, it amounts to $3.2\sigma$ for either $f_{\rm sky} = 2\times 0.44$ and $f_{\rm sky} = 2\times 0.146$, and the significance increases with $\ell_{\rm max}$ (e.g. it is in between 4 and 5$\sigma$ in both cases at $\ell_{\rm max}\approx 2500$). Actually such an asymmetry in the power spectrum constitutes by itself a self-consistency check on the measurement of $\beta$ given in~\cite{Aghanim:2013suk}, as already proposed initially by \cite{Burles:2006xf}. Given this non-negligible result it is thus natural to wonder whether this could account for the asymmetries measured by Planck at $2<\ell<1500$, or at least for a substantial fraction of it.  Looking more closely at the results of \cite{Ade:2013nlj}, v1 sect.~5.5.1, we can see that Planck detected an overall preferred direction with a significance quantified by the fraction of simulations that have smaller clustering of the dipole directions than the data: such number is of about 11/500, 4/500 or less than 1/500, depending on the foreground cleaning methods, which correponds to a $97\%-99\%$ C.L. When looking at our Figs.~\ref{asymmetryell}  and~\ref{asymmetryell2} at $\ell=1500$ we can see that a $1.5-2\sigma$ effect could be easily achieved and it could therefore happen that, once the boost is subtracted, the Planck anomalies would become much less significant or in any case considerably change. Note that in a more recent v2 of \cite{Ade:2013nlj} the claim of an asymmetry at $\ell>600$ has now in fact disappeared\footnote{
The v2 of \cite{Ade:2013nlj} has appeared in December 2013, several months after the first arxiv
version of the present paper.} after removal of the boost, in agreement with the findings of the present paper.

There is also some small effect already at $\ell_{max}\simeq 600$, which should be compared with the asymmetries seen in the  WMAP experiment at $2<\ell<600$ by~\cite{Hansen:2008ym,Hoftuft:2009rq,Axelsson:2013mva}, see also ~\cite{Bennett:2010jb}.
In this case the claimed significance was around $3\sigma$~\cite{Hansen:2008ym,Hoftuft:2009rq,Axelsson:2013mva} and again it is important to reassess once the boost is subtracted, since even a small change in significance could be relevant.

It is also interesting to note that the preferred direction of WMAP in ~\cite{Hansen:2008ym,Hoftuft:2009rq,Axelsson:2013mva}  does not quite point in the same direction as the dipole. 
For Planck the overall direction of the asymmetry is still close to WMAP but it is interestingly a bit shifted towards the dipole direction. In fact, when looking at  Fig.~27 of  v1 of \cite{Ade:2013nlj}, which shows the preferred direction when analyzing the data in $100$--$\ell$ bins, we can see that most bins, especially at low $\ell$, point towards the same direction of WMAP but there are some bins which point instead to the dipole direction, especially at large $\ell$. This seems consistent with our findings that at least a sizable fraction of the asymmetry  may be due to our motion in the direction of the dipole, especially at large $\ell$.  Also in  this case, in the more recent v2 of \cite{Ade:2013nlj} the figure has now in fact changed after removal of the boost, and basically the bins clustered along the dipole direction have moved away, in agreement with the finding of the present paper. Note instead that asymmetries and anomalies at very large scales $\ell\lesssim 60$~\cite{Hansen:2008ym,Hoftuft:2009rq} cannot be accounted for by a boost alone. It is reasonable therefore to consider the possibility  that there may be some intrinsic large scale asymmetry pointing to a different direction which adds up with our boost effect. This hypothesis deserves a more thorough analysis of the real data, going also to $\ell>1500$, where we expect the boost-induced asymmetry to become increasingly relevant. It also crucially stresses the need of analyzing the WMAP and especially Planck data by first removing the aberration and Doppler effects (from the whole map, not only the dipole) and then looking for the true significance of eventual residual anomalies.


Fig.~\ref{asymmetryell} (bottom panel) shows the result in the case in which the hemispheres are not aligned with the dipole. This was achieved by boosting the map along the North--South axis and then rotating it with HEALPix along the actual dipole direction $[(l,b)=(264,48)]$ in galactic coordinates] as measured by the $\ell=1$ multipole of the CMB by WMAP and then performing the cut around the North and South directions. The results we find for the asymmetry parameter $\delta A/A$ is of lower size and significance, as is to be expected since the North-South direction now is not anymore the one which maximizes the asymmetry. In fact, in this particular case $\overline{\beta}$ goes down from 0.56 to 0.42 $\beta$. Finally Fig.~\ref{asymmetryell2} focuses on the case of smaller discs: to two antipodal $90^\circ$-diameter discs. This allows a direct comparison with v1 of~\cite{Ade:2013nlj}.   Note that in this case the size of the effect  is larger, because we are selecting the regions of the sky most affected by a boost (smaller discs around the poles).


We analyze as a relevant example the case of a specific experiment with a different shape, namely ACT, in the next section.

\section{Computing the Bias on ACT Spectrum}\label{sec:act}

\begin{figure}[t]
\begin{minipage}[c]{0.48\linewidth}
    \centering
    \includegraphics[width=\textwidth]{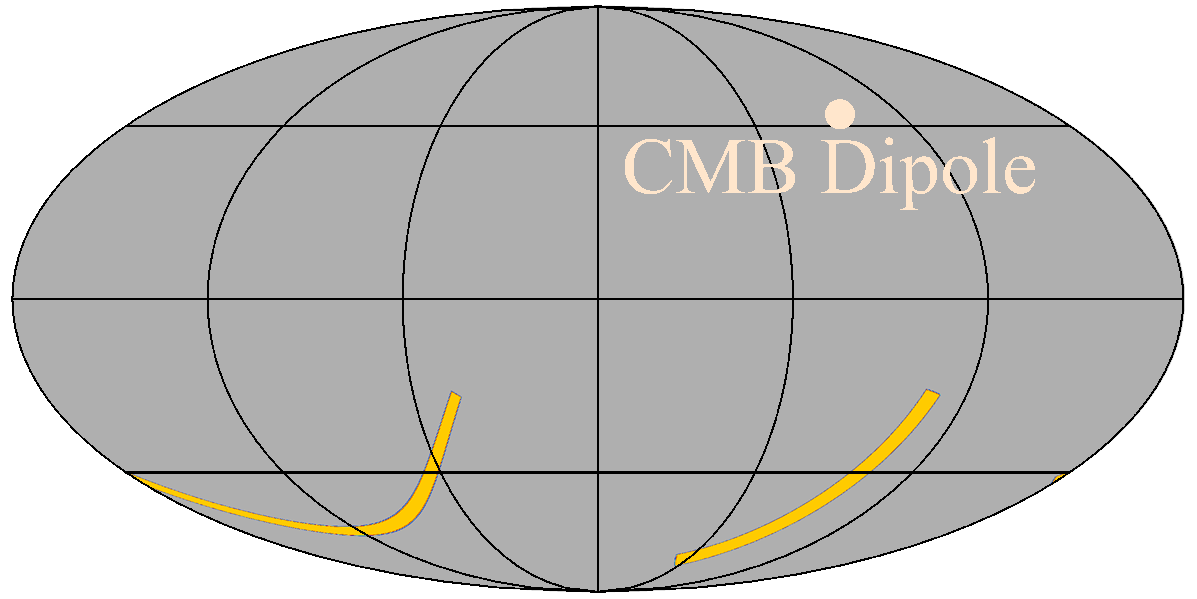}
    \includegraphics[width=\textwidth]{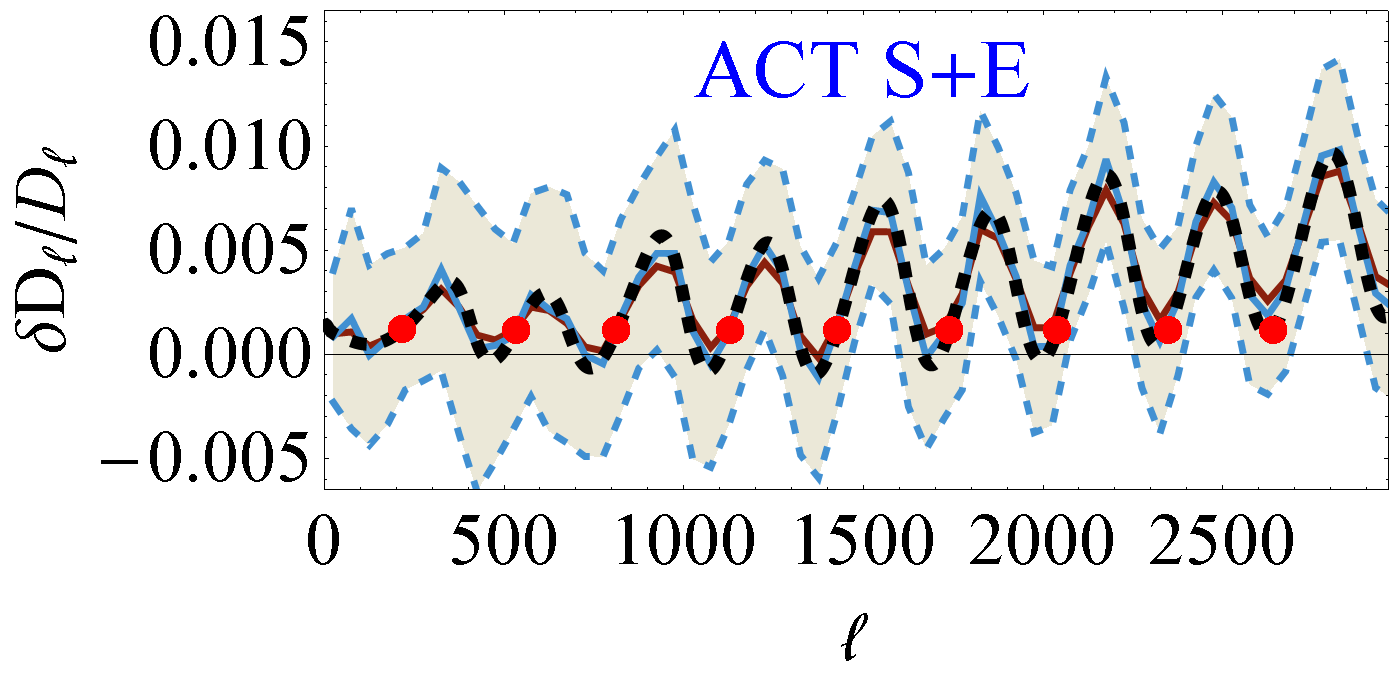}
\end{minipage}
\begin{minipage}[c]{0.52\linewidth}
    \includegraphics[width=\textwidth]{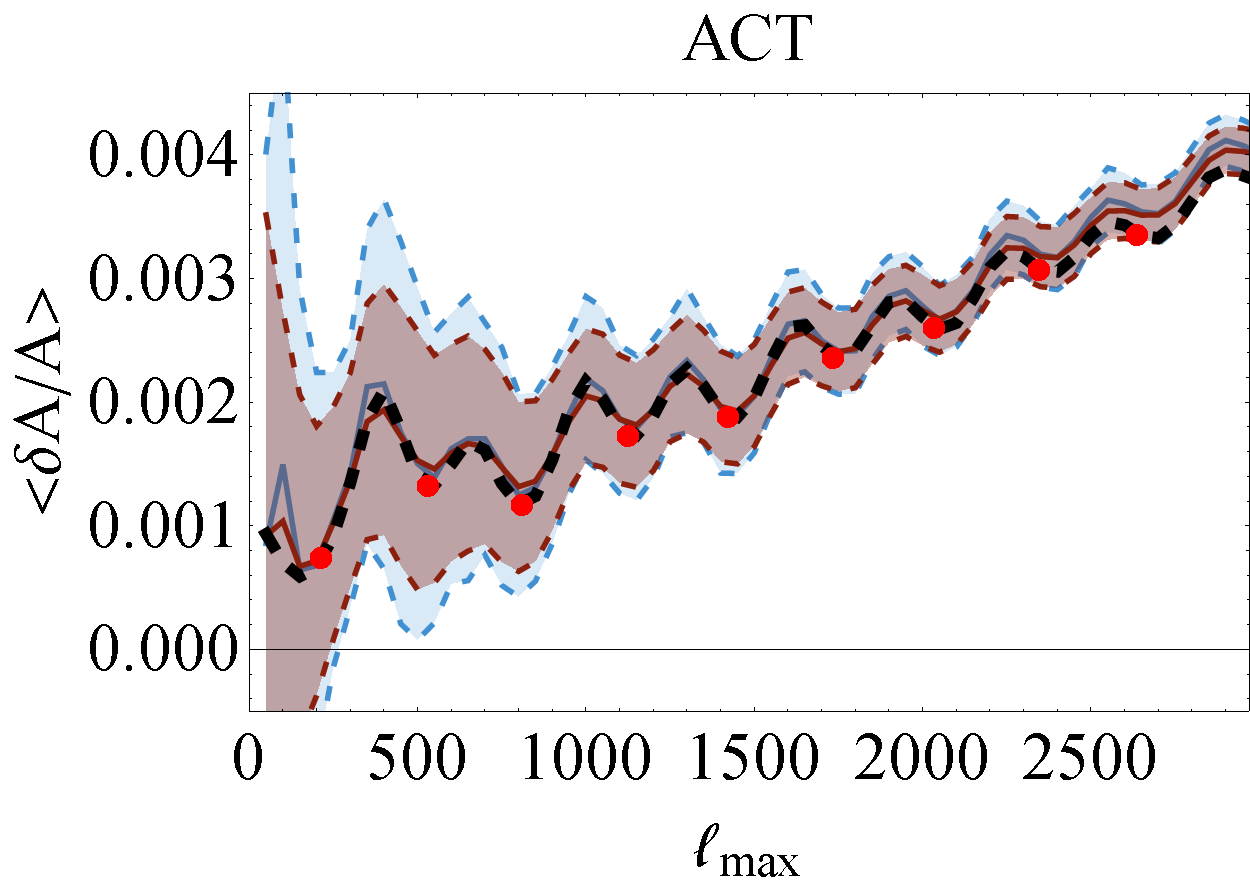}
\end{minipage}
    \caption{Results for the ACT sky-cut. \emph{[Top-left]}: the observed regions, which have a combined $f_{\rm sky} = 0.0143$. \emph{[Bottom-left]}: similar to Fig.~\ref{compare5}, but comparing boosted and un-boosted $D_\ell$, as in Eq.~\eqref{eq:dAoverAestACT}. Here the light-blue curve and region represent the reconstructed $D_\ell$ obtained with the mode-mode coupling matrix and associated $1\sigma$ band; the brown curve is obtained with the raw pseudo-$D_\ell$. \emph{[Right]}: similar to Fig.~\ref{asymmetryell}, for the ACT $D_\ell$'s in the bottom-left plot. Note that the mode-mode coupling matrix only produces significant corrections for low-$\ell$, where the variance is better estimated using better knowledge of the mask.}
    \label{fig:ACT}
\end{figure}

In this section we analyze the bias given by Doppler and Aberration for the ACT~\cite{Sievers:2013wk} experiment, which analyzes two stripes of the sky as shown in Fig.~\ref{fig:ACT}, by using simulations in the simplified case of zero experimental noise. Since the stripes only cover a small region of the sky we ran more simulations (to wit 100, compared to 50 in the previous cases) and we investigated the corrections to our estimate by inverting the mode-mode coupling matrix. All other formulas are unchanged except that we show the boosted spectra $D^B_{\ell}$ minus the unboosted $D_{\ell}$ in Fig.~\ref{fig:ACT}.
The analytic approximation is also similar, apart from a factor of $2$:
\begin{equation}\label{eq:dAoverAestACT}
    \frac{\delta D_\ell}{D_\ell}  \equiv \frac{D_{\ell}-D^B_\ell}{D^{\rm th}_\ell} \;\simeq\; -2 \overline{\beta} \,-\,  \overline{\beta} \, \ell \, \bigg(1-\frac{D_{\ell+1}^{\rm th}}{D_{\ell}^{\rm th}}\bigg)\,,
\end{equation}
and where $\overline{\beta}$ is the average eq. (\ref{averagebeta}) over the two ACT stripes, which gives $\overline{\beta} \simeq -0.51 \beta$.
Comparing with \cite{Jeong:2013sxy} we find good agreement, except that in that paper there is only aberration, while including also Doppler gives another important $2\overline{\beta}$ underestimation bias at all $\ell$, which corresponds to a $0.1\%$ effect. Note that the Doppler effect, being a constant, does not affect the position of the peaks, but just the overall amplitude. The position of the peaks, as already noted by~\cite{Burles:2006xf}, is changed by an amount $\delta\ell / \ell = \overline{\beta}$, which for ACT gives a $0.5\%$ effect, to be compared with the present experimental accuracy, which is of the same order, both considering ACT or Planck. The position of the peaks in current experiments has the smallest error among the relevant cosmological quantities and so it can be affected by the boost effect.  On other parameters  the present precision is instead worse, of order $1\%-2\%$. We leave for future study a full analysis on the effects due to the boost. As it can be seen from Fig. \ref{fig:ACT}, the boost corrections consist in an oscillation which is also growing slowly with $\ell$ and therefore this can affect several parameters, including the overall amplitude, the spectral index and $\Omega_b$ and $\Omega_m$.

We also show in Fig.~\ref{fig:ACT} the bias expected on an amplitude parameter (the usual $\delta A/A$) for such experiment, which gives about $(0.40 \pm 0.03)\%$ at $\ell\simeq 3000$. Note that while the Doppler effect is a constant bias, the aberration effect has some cosmic variance since it depends on the specific realization of the spectrum. Nonetheless the variance is small, and so we conclude that ACT, being in the southern hemisphere, has a bias which corresponds to a  \emph{lack} of overall power of about $0.4\%$.

We also note that for the two stripes of ACT combined the corrections due to the mode-mode coupling matrix are very small, except from the variance at low-$\ell$. Using the $\tilde{C}_\ell$'s one underestimates the variance for $\ell < 1000$, but this becomes irrelevant at higher-$\ell$.

\label{conclusion}
\section{Conclusions}
In this work we have considered the effect of our peculiar velocity on the CMB power spectra in regions in opposite directions of the sky. As in \cite{Catena:2012hq}, we have applied a boost transformation (\ref{boost}) to this frame directly in pixel space on simulated maps, rather than on the $a_{\ell m}$'s, which yields a direct test (although a similar result could be obtained with the fitting functions in~\cite{Notari:2011sb}).
We find that  aberration and Doppler effects induce a directional dependence of the amplitude of the power spectrum in opposite directions which we estimate to be equal to $(0.8 \pm 0.1)\%$ for half sky cuts when summing up to $\ell\lesssim 2500$
and is therefore highly significant. Even if the boost is made along the measured dipole direction $[(l,b)=(264,48)]$, the difference in amplitude between the north and south galactic hemispheres is still very significant, to wit $(0.58 \pm 0.10)\%$. We claim this could be important to understand  the Hemispherical asymmetry found by Planck in the version 1 of ~\cite{Ade:2013nlj}; at least it must account for an important fraction of it. It may have also some small impact on previous detections in the WMAP data at $2<\ell<600$ \cite{Axelsson:2013mva,Hansen:2008ym,Hoftuft:2009rq}, but by itself it cannot explain an anomaly localized only at low-$\ell$ (say, $\ell\leq 60$ as in \cite{Eriksen:2003db,Eriksen:2007pc}). Note that the direction of maximal asymmetry detected in v1 of ~\cite{Ade:2013nlj} and \cite{Axelsson:2013mva,Hansen:2008ym,Hoftuft:2009rq} does not coincide with the dipole direction, but some of the bins used for the analysis in v1 of~\cite{Ade:2013nlj} do indeed point towards the dipole direction: this suggests that the asymmetries reported by Planck are likely to be a mixture of a boost effect, showing up especially at high $\ell$, with some residual asymmetry mostly relevant at low $\ell$. The new versions v2 and v3 of the Planck paper ~\cite{Ade:2013nlj} find now in fact that the asymmetry is significant only up to $\ell\lesssim 600$ after removing the boost effect. Moreover the bins of the previous v1 analysis have consistently moved away from the dipole direction.
Note that we are relying on the standard assumption that our velocity is given by the CMB dipole; if that is not the case there might be a discrepancy between the dipole and aberrated directions.  In any case our findings clearly indicate that before looking for preferred directions in the real data at high multipoles one should properly deboost the map going to the CMB rest frame to avoid spurious detection of anomalies.
We have also shown that experiments which observe small fractions of the sky, such as SPT or ACT, can suffer of a bias up to about $1\%$  on the power spectrum. More specifically for an experiment like ACT the position of the CMB peaks can shift of about $\overline{\beta} \simeq 0.5\%$ \cite{Burles:2006xf, Jeong:2013sxy} and we also showed that it suffers from a bias in the power spectrum of about $1\%$  (which has an oscillatory shape due to the peak structure of the CMB) which corresponds to a {\it lack} of power in the overall amplitude of $-0.4\%$ at $\ell\sim 3000$. Presumably the bias could be of similar size on other Cosmological parameters. A similar bias affects any small-$f_{\rm sky}$ experiment and it grows for better angular resolutions.

\acknowledgments
Some of the results in this paper have been derived using the HEALPix package \cite{Gorski:2004by}.
We thank  Carlos Hernandez-Monteagudo, Massimiliano Lattanzi, Michele Liguori, Jordi Miralda, Paolo Natoli and Jorge Nore\~ na  for useful discussions and comments. MQ is grateful to Brazilian research agencies CNPq and FAPERJ for support. AN is supported by the grants EC FPA2010-20807-C02-02, AGAUR 2009-SGR-168. RC acknowledges partial support from the European Union FP7 ITN INVISIBLES (Marie Curie Actions, PITN-GA-2011-289442).

\subsection*{Note Added} After submission to the arXiv of the first version of this paper similar results on the bias on small-scale experiments have been analyzed by~\cite{Jeong:2013sxy}, showing that an ${\cal}O(1\%)$ bias is present in the power spectrum, consistently with our findings (although they neglect the Doppler effect, which is non-negligible). Also, another paper appeared~\cite{Flender:2013jja} concluding that a doppler boost was a significant source of hemispherical asymmetry in Planck XXIII's original results. Finally, as we have stressed already in the main text, the Planck paper has changed its conclusions in its v2 and v3 which appeared on December 2013 and January 2014, several months after our arxiv version: the claim of an asymmetry is now limited only up to $\ell<600$ after removal of the boost effect and the bins of their analysis have moved away from the dipole direction, consistently with the predictions of the present paper.

\appendix
\section{Including aberration and Doppler in HEALPix}

In this appendix we provide details regarding the modifications made to the HEALPix code to include Doppler and aberration effects. The aim is to clarify the procedure used in this work to generate from a fiducial angular power spectrum a temperature map which includes Doppler and aberration, {\it i.e.} as seen by an observer which is moving with respect to the CMB rest frame.

Our approach is based on a modification of the \emph{synfast} program (and of related subroutines). We start from the following observations. Neglecting Doppler and aberration, the CMB temperature at the center of a generic pixel $p$ assumes the value
\begin{equation}
T(\theta_p,\phi_p) = \sum_{\ell m} a_{\ell m} Y_{\ell m}(\theta_p,\phi_p)
\end{equation}
where the angles $(\theta_p,\phi_p)$ identify the position of the center of the pixel $p$ and the coefficients $a_{\ell m}$ have been sampled from the assumed fiducial angular power spectrum. Including Doppler and aberration the previous equation is modified as follows
\begin{equation}
T(\theta_p,\phi_p) = \Big[1+\beta\cos\big(f(\theta_p)\big)\Big] \sum_{\ell m} a_{\ell m} Y_{\ell m}\big(f(\theta_p),\phi_p\big)
\label{boost2}
\end{equation}
where the function $f(\theta_p)$ is given by\footnote{To compare these equations with Eq.~(\ref{boost}) one has to impose: $T(\theta_p,\phi_p) = T^{\prime}(\hat{\mathbf{n}}')$, $\cos\theta = \cos\big(f(\theta_p)\big)$ and finally $\cos\theta^{\prime} = \cos\theta_p$.}

\begin{figure}[t]
\begin{center}
\includegraphics[width=.47\columnwidth]{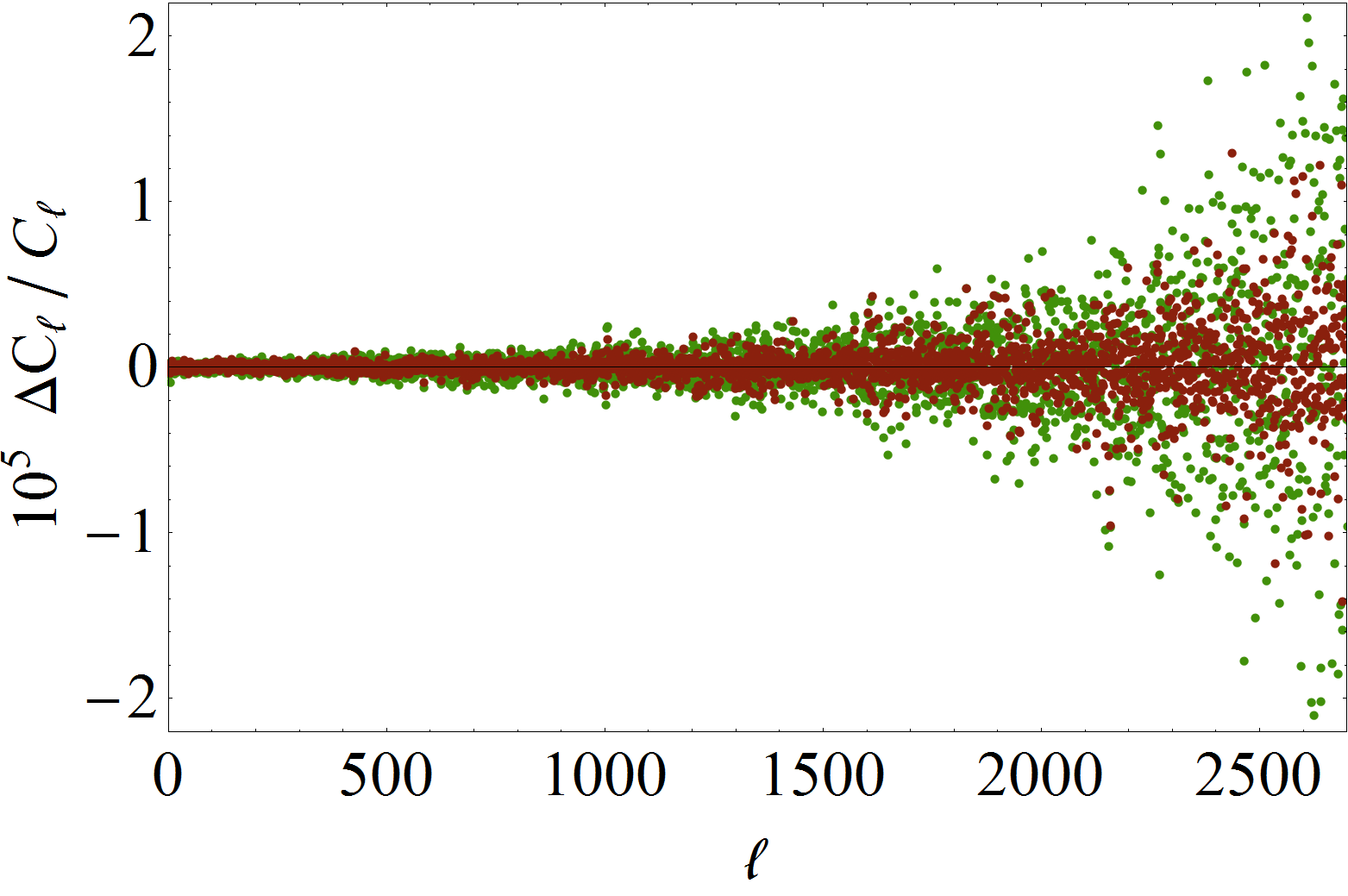}\quad
\includegraphics[width=.49\columnwidth]{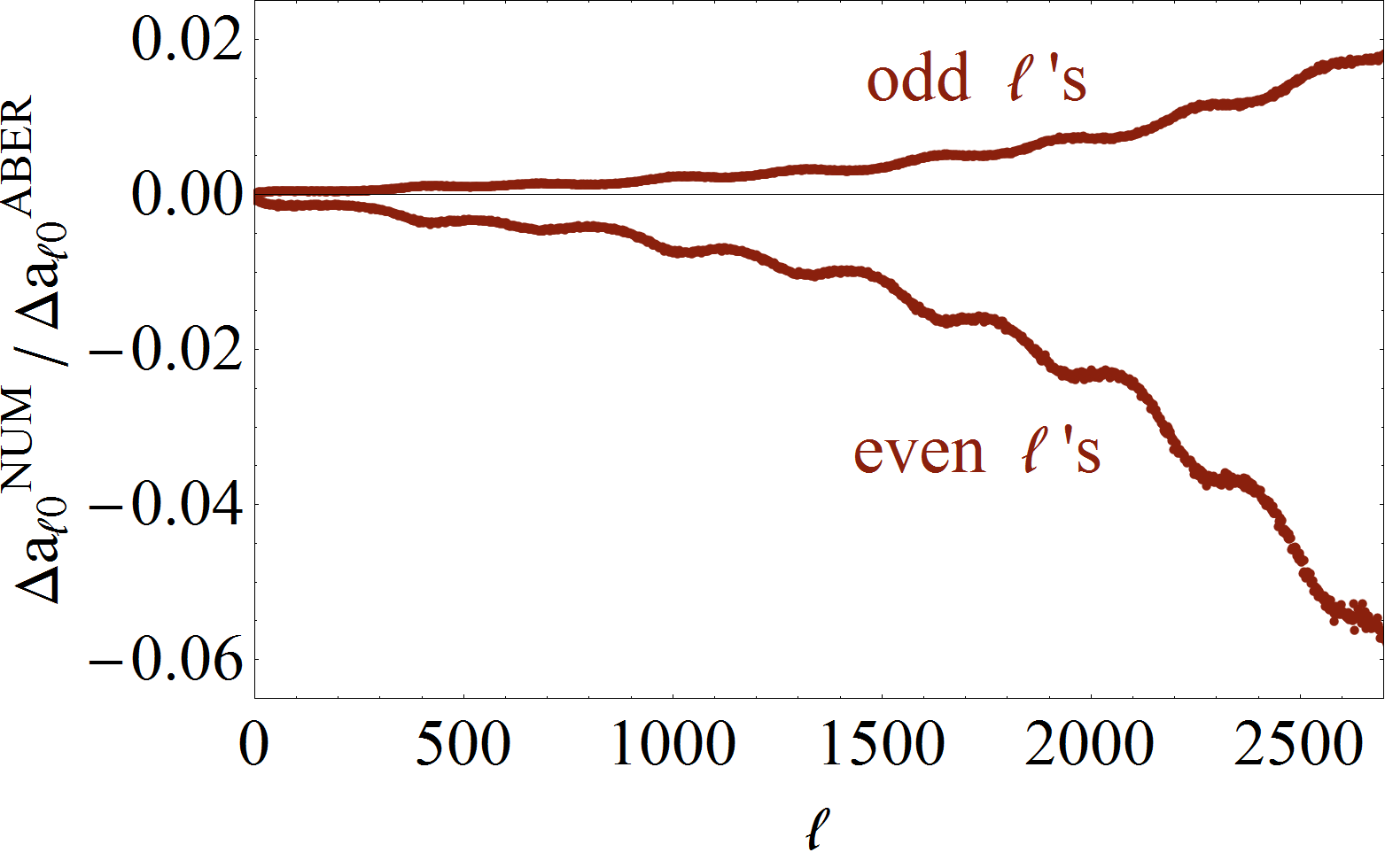}
\end{center}
    \caption{Analysis on the numerical accuracy of our code. \emph{[Left]}:
    Relative difference, $\Delta D_\ell/D_\ell$, between a set of $D_\ell$ simulated by \emph{synfast} and a different set of $D_\ell$ obtained from the former boosting the associated temperature map with $\beta$ and then de-boosting the same temperature map (with $-\beta$) as a function of $\ell$. The brown curve corresponds to our program while the green curve is associated with the original HEALPix code (boosting and de-boosting with $\beta = 0$).
    \emph{[Right]}: Ratio between the numerical errors and the aberration effect for the particular case of $m=0$. Depicted is Eq.~\eqref{eq:delta-alm}, to wit a fraction in which the numerator is the difference between a set of $a_{\ell 0}$ (simulated by \emph{synfast}) and a different set of $a_{\ell m}$ obtained after boosting and then deboosting in Healpix (see text) and the denominator is $\Delta a_{\ell 0}^{\rm Aberration}$, given by Eq.~\eqref{eq:delta-al0-aber}. Note that although the numerical errors do increase with $\ell$ and that there are biases depending on the parity of $\ell$, they remain subdominant.}
\label{Delta}
\end{figure}

\begin{equation}
f(\theta_p) = \arccos\left( \frac{\cos\theta_p-\beta}{1-\beta\cos\theta_p} \right)
\end{equation}
and $\beta=1.23\times 10^{-3}$. Notice that in general the angles $\big(f(\theta_p),\phi_p\big)$ do not identify the center of a pixel, nevertheless $Y_{\ell m}\big(f(\theta_p),\phi_p\big)$ can be calculated exactly since the spherical harmonics and the function $f(\theta_p)$ are known at any point. Implementing the function $f(\theta_p)$ in the \emph{get\_pixel\_layout} subroutine and including the Doppler pre-factor $\left[1+\beta\cos(f(\theta_p))\right]$ in the subroutine \emph{alm2map} we where therefore able to evaluate exactly Eq.~(\ref{boost2}) at the centers of the HEALPix pixels for any assumed $N_{side}$ and fiducial angular power spectrum. In fact, the left hand side of this equation gives the desired CMB temperature map in the boosted frame evaluated at $(\theta_p,\phi_p)$, where the angles $(\theta_p,\phi_p)$ identity the centers of the HEALPix pixels.


In Fig.~\ref{Delta} (left-hand side) we show the accuracy of this procedure plotting the relative difference between a set of $D_\ell$ simulated by \emph{synfast} and a different set of $D_\ell$ obtained from the former boosting the associated temperature map (with $\beta=1.23\times 10^{-3}$) and then de-boosting the same temperature map (with $\beta=-1.23\times 10^{-3}$). This requires to run the sequence of routines: \emph{synfast}($\beta$), \emph{anafast}, \emph{synfast}($-\beta$), \emph{anafast}. The two sets of $D_\ell$ should be identical and indeed we find a relative difference between them which is always much less then $10^{-4}$. In the same figure we also show what we find when using the original HEALPix and following the same sequence of routines. Our procedure introduces a numerical error which is only mildly larger.

As an additional test of the accuracy within which we were able to implement aberration and Doppler effect in the HEALPix code, we also compared our benchmark  $a_{\ell m}$ with the ones obtained after applying the sequence of routines:  \emph{synfast}($\beta$), \emph{anafast}, \emph{synfast}($-\beta$), \emph{anafast} (with $\beta=1.23\times 10^{-3}$). The result of this calculation is also shown in Fig.~\ref{Delta} (in the right-hand side), where we plot the difference between the two sets of $a_{\ell m}$ mentioned above divided by the expected average difference in the $a_{\ell m}$'s due to a single boost. In more detail, we compute
\begin{equation}\label{eq:delta-alm}
    \frac{\Delta a_{\ell m}^{\rm boost-deboost}}{\Delta a_{\ell m}^{\rm Aberration}}\,,
\end{equation}
where the numerator is an estimate of the numerical errors involved and $\Delta a_{\ell m}^{\rm Aberration} = a_{\ell m}^{\rm Aberrated} - a_{\ell m}^{\rm Primordial}$, which in particular for $\,m=0\,$ and $\,\ell \gtrsim 10\,$ is given by~\cite{Amendola:2010ty}
\begin{equation}\label{eq:delta-al0-aber}
    \Delta a_{\ell 0}^{\rm Aberration} \simeq \frac{1}{2} \beta \ell \Big[ a_{\ell+1,0} - a_{\ell-1,0} \Big] \sim \frac{1}{2} \beta \ell \sqrt{C_\ell}\,.
\end{equation}
As can be seen in Fig.~\ref{Delta}, although the numerical errors show biases depending on the parity of $\ell$ which increase at large $\ell$, they remain subdominant in the $\ell$--range here considered. Although it might be interesting to understand better the source of such bias, the important conclusion here is that they can be neglected for the purposes of computing the asymmetry. The asymmetry is directly related to the angular power spectrum, and the results for the $D_\ell$'s corroborate our conclusion regarding the negligible influence of the small biases above. Furthermore, as discussed in Section~\ref{sec:asymmetry}, both the numerical code and analytical approximations (to wit, Eqs.\eqref{eq:dAoverAest} and \eqref{averagebeta}) agree with each other, providing yet another indirect confirmation of the robustness of our results.


\bibliographystyle{JHEP}
\bibliography{asymmetry}

%

\end{document}